\documentclass[fleqn,usenatbib,letters]{mnras}

\usepackage{newtxtext,newtxmath}

\usepackage[T1]{fontenc}

\DeclareRobustCommand{\VAN}[3]{#2}
\let\VANthebibliography\thebibliography
\def\thebibliography{\DeclareRobustCommand{\VAN}[3]{##3}\VANthebibliography}

\usepackage{graphicx}	
\usepackage{amsmath}	
\usepackage{scalerel}
\usepackage{multirow}

\newcommand{\Teff}{\mbox{$T_{\rm eff}$}}

\title[Thin H layer on double-faced white dwarf]{The prototype double-faced white dwarf has a thin hydrogen layer across its entire surface}

\author[A. B\'edard and P.-E. Tremblay]{
Antoine B\'edard\thanks{E-mail: antoine.bedard@warwick.ac.uk}
and Pier-Emmanuel Tremblay
\\
Department of Physics, University of Warwick, Coventry, CV4 7AL, UK
}

\date{Accepted 2025 January 27. Received 2025 January 24; in original form 2024 December 18}

\pubyear{\the\year{}}

\begin{document}
\label{firstpage}
\pagerange{\pageref{firstpage}--\pageref{lastpage}}
\maketitle

\begin{abstract}
Some white dwarfs undergo significant changes in atmospheric composition owing to the diffusion and mixing of residual hydrogen in a helium-rich envelope. Of particular interest are a few objects exhibiting hydrogen and helium line variations modulated by rotation, revealing surface composition inhomogeneities. Recently, the hot ultramassive white dwarf ZTF\,J203349.80+322901.1 emerged as the most extreme such specimen, with hydrogen and helium lines successively appearing and vanishing in anti-phase, suggesting a peculiar double-faced configuration. However, standard atmosphere models fail to reproduce the observed spectrum at all rotation phases, hampering further interpretation. Here, we perform a new analysis of ZTF\,J203349.80+322901.1 using stratified atmosphere models, where hydrogen floats above helium, and obtain excellent fits to the phase-resolved spectra. Our results imply that an extremely thin hydrogen layer covers the entire surface but varies from optically thick to optically thin across the surface, thus producing the observed spectral variations. We present new envelope models indicating that the hydrogen layer arises from a delicate interplay between diffusion and convection. We discuss possible explanations for the surface layer asymmetry, including an asymmetric magnetic field and a non-uniform internal hydrogen distribution. Finally, we highlight implications for expanding and understanding the emerging class of inhomogeneous white dwarfs.
\end{abstract}




\section{Introduction}
\label{sec:intro}

White dwarfs are dense stellar remnants in which the strong inward gravitational pull efficiently purifies the outer layers of heavy elements. In principle, this gravitational settling process should produce an atmosphere dominated by the lightest chemical species present in the star --- hydrogen in most cases, or helium if the hydrogen content was drastically reduced in a previous evolutionary phase. However, a wealth of evidence reveals that the surface composition of white dwarfs can change over time owing to various element transport mechanisms, a complex and far-reaching phenomenon known as spectral evolution (see \citealt{bedard2024b} for a review).

The case of hydrogen-deficient white dwarfs is particularly remarkable. As these remnants enter their terminal cooling stage, they initially exhibit a helium-rich atmosphere, but most of them develop a hydrogen-rich atmosphere before reaching an effective temperature $\Teff \simeq 40\,000$\,K, changing their spectral type from DO to DA \citep{bedard2020}. These objects then revert back to a helium-rich atmosphere after cooling below $\Teff \simeq 20\,000$\,K, causing another change of spectral type from DA to DB(A) or DC \citep{genest-beaulieu2019b, cunningham2020, lopez-sanjuan2022, torres2023, vincent2024, kilic2024b}. 
These transformations are understood to arise from the presence and transport of a small amount of residual hydrogen in the helium-dominated envelope. The hydrogen is initially diluted within the envelope and thus invisible, but then gradually diffuses upward and eventually forms a thin layer at the surface, leading to the DO-to-DA transition \citep{unglaub2000, althaus2005b, althaus2020a, bedard2022a, bedard2023}. As the star cools further, convection develops in the envelope and ultimately mixes the thin hydrogen layer back into the large helium reservoir, resulting in the DA-to-DB(A)/DC transition \citep{macdonald1991, chen2011, rolland2018, rolland2020, cunningham2020, bedard2022a, bedard2023}. Such an evolution implies that the total mass of hydrogen contained in the star is only $\sim$10$^{-12}$--10$^{-8}$\,$M_*$ (where $M_*$ is the stellar mass), many orders of magnitude smaller than in standard hydrogen-rich white dwarfs, $\sim$10$^{-4}$\,$M_*$ \citep{renedo2010, bedard2023}.

The transformations described above involve evolutionary phases where both hydrogen and helium are simultaneously present in the atmosphere and thus detectable in the spectrum. At high temperature, the gradual build-up of the surface hydrogen layer through diffusion gives rise to the so-called stratified atmosphere white dwarfs. In these objects, the nascent hydrogen layer is still thin enough ($\sim$10$^{-19}$--10$^{-15}$\,$M_*$) that the underlying helium remains visible. This phase is necessarily short-lived given the continuous influx of hydrogen, and therefore only about 30 stratified atmosphere white dwarfs are currently known, mostly between $\Teff \simeq 55\,000$\,K and $35\,000$\,K \citep{manseau2016, bedard2020}. At lower temperature, the dilution of the hydrogen layer by convection results in a helium-dominated atmosphere with a uniformly mixed trace of hydrogen, appearing as a DBA white dwarf. These mixed atmospheres, mostly found below $\Teff \simeq 20\,000$\,K, are much more common than their hotter stratified counterparts: over 1000 DBA stars are currently known, and they may represent up to 75 per cent of the helium-rich population in this temperature range \citep{bergeron2011, koester2015b, gentile-fusillo2017, rolland2018, genest-beaulieu2019b, cukanovaite2021}. In general, the chemical structure, either stratified or mixed, can be unambiguously determined via model atmosphere analysis as it leaves a clear imprint in the optical spectrum, with stratified white dwarfs exhibiting shallower lines \citep{manseau2016, bedard2020}.

Interestingly, a small number of white dwarfs show recurrent variations in the strength of their hydrogen and helium lines on timescales of minutes to years. This peculiar group includes the long-known objects Feige\,7 \citep{achilleos1992}, G104--27 \citep{kidder1992}, PG\,1210+533 \citep{bergeron1994, gianninas2010}, HS\,0209+0832 \citep{heber1997, wolff2000}, PG\,0853+164 \citep{wesemael2001}, and GD\,323 \citep{koester1994, pereira2005}. These were recently joined by the newest members ZTF\,J203349.80+322901.1 (hereafter ZTF\,J2033; \citealt{caiazzo2023}), SDSS\,J091016.43+210554.2 \citep{moss2024}, GALEX\,J071816.4+373139 \citep{cheng2024}, and SDSS\,J084716.21+484220.4 \citep{moss2025}. In most cases, the variability has been linked to the rotation of the star, thus signalling the presence of composition inhomogeneities at the surface. Among these, ZTF\,J2033, discovered and nicknamed Janus\footnote{We refrain from referring to ZTF\,J203349.80+322901.1 as Janus, as it is not an official IAU designation. In fact, the name Janus is already formally assigned to one of Saturn's moons \citep{west1983}.} by \citet{caiazzo2023}, stands out as the most extreme specimen, as this hot ultramassive white dwarf successively shows a pure DA and pure DB spectrum over its rotation period of $\simeq$15 minutes. These variations suggest an unusual double-faced configuration, with an atmosphere dominated by hydrogen on one side and by helium on the other. \citet{caiazzo2023} tentatively explained this as the result of a spectral evolution event in the presence of a weak asymmetric magnetic field, which would affect diffusion and/or convection more strongly on one side than the other. However, further interpretation is hampered by the inability of standard atmosphere models (with either pure-hydrogen, pure-helium, or uniformly mixed compositions) to reproduce the spectrum at all rotation phases, the observed lines appearing anomalously shallow.

The latter discrepancy is a telltale signature of a stratified atmosphere where a thin hydrogen layer floats atop a helium mantle. In this Letter, we perform a new analysis of ZTF\,J2033 using stratified atmosphere models and demonstrate that these models provide an excellent fit to the observations reported by \citet{caiazzo2023}. This leads us to conclude that the entire surface of ZTF\,J2033 is covered by a hydrogen layer of varying thickness, implying that the physical difference between the DA and DB faces is less drastic than previously believed. Section \ref{sec:atmo} details our improved analysis of the available time-resolved spectroscopy and photometry. Section \ref{sec:env} presents new envelope models aimed at exploring the interplay between diffusion and convection in the outer layers. Finally, Section \ref{sec:conclu} reviews the possible astrophysical interpretations in light of our results and concludes with future prospects.

\section{Spectroscopic and photometric analysis}
\label{sec:atmo}

The double-faced character of ZTF\,J2033 manifests itself through rotation-modulated variations in both overall brightness and spectral morphology. \citet{caiazzo2023} obtained time-resolved photometry in the UltraViolet and Optical Telescope (UVOT) UVW2 band \citep{roming2005} and the HiPERCAM $u_{\mathrm{s}} g_{\mathrm{s}} r_{\mathrm{s}} i_{\mathrm{s}} z_{\mathrm{s}}$ bands \citep{dhillon2021}. They measured sinusoidal light curves with a period of 14.97 min and peak-to-peak amplitude variations ranging from $0.12 \pm 0.02$\,mag in the $z_{\mathrm{s}}$ band to $0.46 \pm 0.08$\,mag in the UVW2 band. They also obtained time-resolved optical spectroscopy using the Low-Resolution Imaging Spectrometer (LRIS) on the Keck I Telescope \citep{oke1995}. They found that ZTF\,J2033 successively shows only hydrogen lines and only helium lines at the phases of maximum and minimum brightness, respectively, with a smooth transition in between. \citet{caiazzo2023} estimated the stellar parameters by fitting atmosphere models to the observed photometry at maximum and minimum brightness, allowing for distinct effective temperatures. Based on the spectroscopic data, they assumed a pure-hydrogen composition for the brighter DA face and a helium-rich composition (allowing for uniformly mixed traces of hydrogen) for the fainter DB face. They derived $\Teff = 34\,900_{\scaleto{-1500}{4pt}}^{\scaleto{+1300}{4pt}}$\,K for the DA face and $\Teff = 36\,700_{\scaleto{-1600}{4pt}}^{\scaleto{+1300}{4pt}}$\,K for the DB face, hinting that the net flux variation may arise from both a difference in composition and a difference in temperature. They also inferred a stellar mass $M_* = 1.21 \pm 0.06\,M_{\odot}$ assuming an oxygen--neon core. Nevertheless, the model spectra predicted from their photometric solution did not match the observed spectra, the observed hydrogen and helium lines appearing systematically weaker than expected. They noted that these shallow features seemingly indicate a much higher temperature ($\Teff \simeq 50\,000$\,K), which is however ruled out by the spectral energy distribution and the absence of the \ion{He}{ii}\,$\lambda$4686 line in the spectrum.

\begin{figure*}
\includegraphics[width=\textwidth,clip=true,trim=2.3cm 17.9cm 2.3cm 3.6cm]{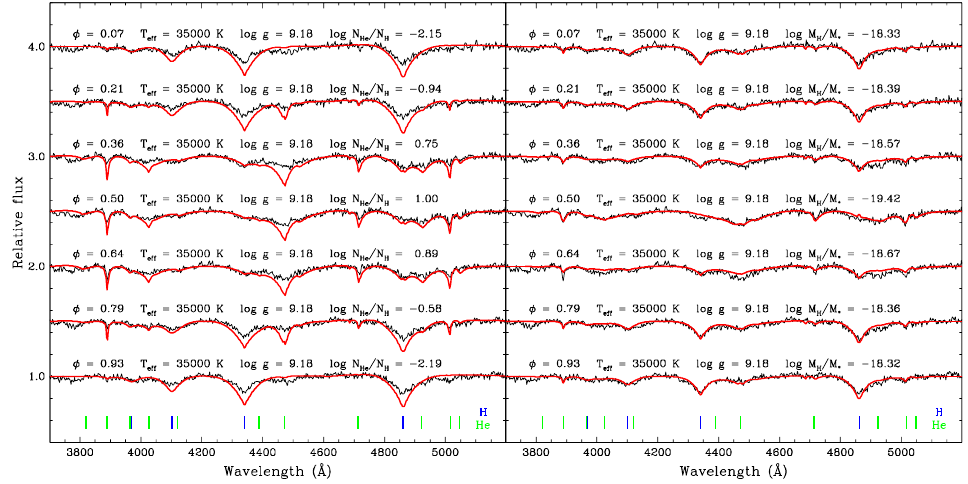}
\caption{Best fits to the phase-resolved spectra of ZTF\,J2033 using two sets of atmosphere models: standard models where hydrogen and helium are uniformly mixed (left panel), and stratified models where hydrogen floats above helium in diffusive equilibrium (right panel). The observed and synthetic spectra are shown as black and red curves, respectively. The spectra are normalised and shifted vertically from each other by 0.5 for clarity. Each spectrum is labelled with the rotation phase, effective temperature, surface gravity, and atmospheric composition (either the helium-to-hydrogen number abundance for the mixed models or the hydrogen layer mass for the stratified models). In all fits, the temperature and gravity were held fixed while the composition parameter was allowed to vary. Blue and green ticks mark the location of hydrogen and helium lines, respectively.}
\label{fig:spec}
\end{figure*}

As similar discrepancies seen in other hot white dwarfs have been resolved by invoking stratified atmospheres (see fig.\,10 of \citealt{bedard2020}), we set out to reanalyse ZTF\,J2033 under that assumption, first focusing on the spectroscopic observations. We retrieved the LRIS spectra (available at \url{https://github.com/ilac/Janus}) and combined them into seven phase-binned spectra as in fig.\,2 of \citet{caiazzo2023}, with $\phi = 0.0$ and $0.5$ corresponding to the pure DA and pure DB phases, respectively. The normalised phase-binned spectra are shown in Fig.\,\ref{fig:spec}. To connect with previous work, we first fitted these spectra with standard atmosphere models where hydrogen and helium are uniformly mixed. In these models, the atmospheric composition is parametrised as usual by the helium-to-hydrogen number abundance ratio, $N_{\mathrm{He}}/N_{\mathrm{H}}$. The models and fitting algorithm are described in \citet{bedard2020}. For reasons that will appear in due course, we fitted only for the atmospheric composition while fixing $\Teff = 35\,000$\,K and a surface gravity $\log g = 9.18$ at all phases. The best-fitting model spectra and parameters are displayed in the left panel of Fig.\,\ref{fig:spec}. Expectedly, all lines are predicted too strong at all phases, illustrating the discrepancy reported by \citet{caiazzo2023}, who however did not present such formal fits. The inferred helium-to-hydrogen ratio increases between $\phi = 0.0$ and $0.5$ and then decreases between $\phi = 0.5$ and $1.0$, in line with the overall trend in spectral appearance, but the abundance values are essentially meaningless given the poor agreement. Furthermore, we verified that fits where the temperature is allowed to vary (not shown here) yield significantly hotter solutions (closer to $\Teff \simeq 45\,000$\,K), moderately improving the match to some lines but also predicting a strong \ion{He}{ii}\,$\lambda$4686 feature that is not observed.

\begin{figure}
\includegraphics[width=\columnwidth,clip=true,trim=5.3cm 20.3cm 5.5cm 3.5cm]{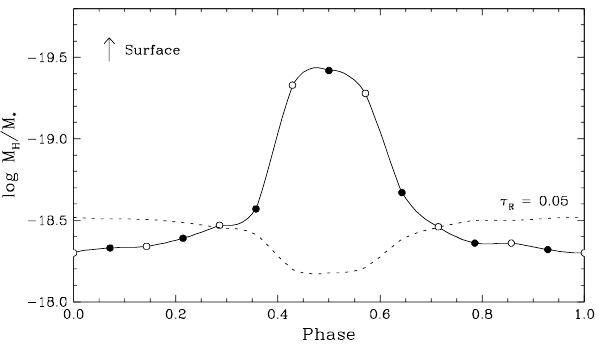}
\caption{Mass of the surface hydrogen layer of ZTF\,J2033 as a function of phase, as inferred from our spectroscopic analysis assuming diffusive equilibrium. The filled points correspond to the results shown in the right panel of Fig.\,\ref{fig:spec}, the empty points denote the results of similar fits using shifted phase bins for better phase coverage, and the solid curve is a cubic interpolation. The dotted curve shows the mass depth where the Rosseland optical depth is 0.05 as a rough indication of the line formation region. Note that the total mass of hydrogen in the envelope may be systematically larger owing to the occurrence of convective mixing (see Section\,\ref{sec:env}).}
\label{fig:layer}
\end{figure}

We then repeated the analysis using the stratified atmosphere models of \citet{bedard2020}, where hydrogen floats above helium in diffusive equilibrium. These models are parametrised by the mass of the surface hydrogen layer, expressed in terms of the total stellar mass, $M_{\mathrm{H}}/M_*$. We again assumed $\Teff = 35\,000$\,K and $\log g = 9.18$ and fitted only for the hydrogen layer mass at a given phase. The best-fitting model spectra and parameters are displayed in the right panel of Fig.\,\ref{fig:spec}. Clearly, the stratified configuration provides a much better fit to the observed spectrum over the full rotation period. Fig.\,\ref{fig:layer} shows the inferred hydrogen layer mass as a function of phase; we find $\log M_{\mathrm{H}}/M_* \simeq -18.3$ at $\phi = 0.0$, monotonically decreasing to $\log M_{\mathrm{H}}/M_* \simeq -19.4$ at $\phi = 0.5$, and then increasing nearly symmetrically until $\phi = 1.0$. Fig.\,\ref{fig:layer} also indicates the mass depth where the Rosseland optical depth $\tau_{\mathrm{R}} = 0.05$, approximately denoting the region where spectral lines are formed. Overall, these results strongly suggest that ZTF\,J2033 has a thin hydrogen layer covering its entire surface. The thickness of this layer varies across the surface, such that it is optical thick on one side and optically thin on the other, thus causing the hydrogen and helium lines to appear and disappear in anti-phase. Furthermore, we note from Fig.\,\ref{fig:layer} that the region where the hydrogen layer is thinner spans less than a full hemisphere.

Although a significant improvement, the stratified fits of Fig.\,\ref{fig:spec} are not perfect, especially around the \ion{He}{i}\,$\lambda$4471 and $\lambda$4922 lines at $\phi = 0.36$ and $0.64$. Our analysis involves two approximations that may account for these residual discrepancies. First, the stratified atmosphere models of \citet{bedard2020} assume that the hydrogen layer mass of a given model is constant over the stellar disk. This implies that our fits measure the average hydrogen mass over the visible stellar disk at a given phase, an approach that is presumably less accurate at $\phi = 0.36$ and $0.64$ where this quantity changes rapidly (Fig.\,\ref{fig:layer}). Second, these models assume that the hydrogen and helium layers are in diffusive equilibrium, meaning that the possible effect of convection is ignored. As we will see in Section \ref{sec:env}, convective mixing in the helium-rich mantle may extend the hydrogen diffusion tail to greater depths, and thus our atmosphere models may not reliably capture the composition profile below the hydrogen-rich layer. We come back to this point in Section\,\ref{sec:env}.

\begin{figure}
\includegraphics[width=\columnwidth,clip=true,trim=5.4cm 20.3cm 5.4cm 3.5cm]{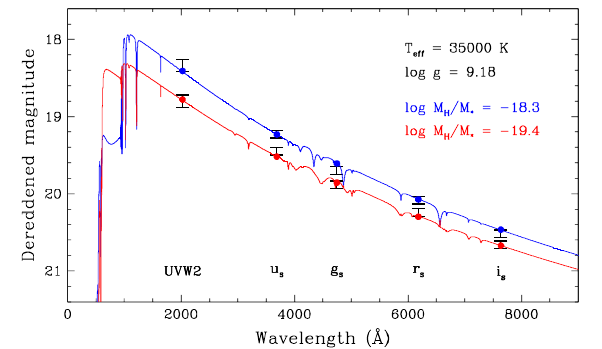}
\caption{Best fit to the dereddened UVOT and HiPERCAM photometry of ZTF\,J2033 at maximum and minimum brightness ($\phi = 0.0$ and $0.5$, coinciding with the pure DA and pure DB phases, respectively) using stratified atmosphere models. The observed magnitudes are shown as black error bars, while the synthetic magnitudes are shown as blue ($\phi = 0.0$) and red ($\phi = 0.5$) points. The full synthetic spectra are displayed as solid curves of the same colours. The labels specify the (phase-independent) effective temperature and surface gravity and the (phase-dependent) hydrogen layer masses. In the fit, the temperature and radius (hence gravity via the mass--radius relation) were allowed to vary while the hydrogen masses were held fixed.}
\label{fig:phot}
\end{figure}

Having established the stratified structure of ZTF\,J2033 using spectroscopy, we then turned to photometry to investigate the brightness variations under this interpretation. We analysed the available UVOT and HiPERCAM photometry at maximum and minimum brightness similarly to \citet{caiazzo2023}, but relying on stratified atmosphere models. We followed the usual approach of fitting for the effective temperature and stellar radius while assuming the distance obtained from the \textit{Gaia} parallax ($D = 408_{\scaleto{-61}{4pt}}^{\scaleto{+85}{4pt}}$\,pc; \citealt{gaia2023a}) and the spectroscopically derived composition ($\log M_{\mathrm{H}}/M_* = -18.3$ and $-19.4$ for the brighter DA and fainter DB sides, respectively). We adopted the simplest hypothesis of a uniform temperature across the surface to test whether it could provide a satisfactory fit within our improved framework. We dereddened the measured magnitudes using $E(B-V) = 0.036$ \citep{lallement2019} and a standard extinction law \citep{cardelli1989}. We computed synthetic magnitudes from our atmosphere models by integrating the monochromatic flux over the appropriate transmission functions and converting these average fluxes to the AB magnitude system. Our best fit is shown in Fig.\,\ref{fig:phot}; we find $\Teff = 35\,000 \pm 1500$\,K and a stellar radius $R_* = 0.0047_{\scaleto{-0.0007}{4pt}}^{\scaleto{+0.0010}{4pt}}\,R_{\odot}$, corresponding to $M_* = 1.24_{\scaleto{-0.07}{4pt}}^{\scaleto{+0.05}{4pt}}\,M_{\odot}$, $\log g = 9.18_{\scaleto{-0.20}{4pt}}^{\scaleto{+0.16}{4pt}}$, and a cooling age $t_{\mathrm{cool}} = 145_{\scaleto{-60}{4pt}}^{\scaleto{+65}{4pt}}$\,Myr based on oxygen--neon core evolutionary models \citep{camisassa2019}. Therefore, we achieve a unique solution that reproduces both the spectroscopic and photometric data, and our analysis is fully self-consistent (justifying the $\Teff$ and $\log g$ values assumed in our spectroscopic fits). Moreover, we find that the brightness variations are entirely accounted for by the varying hydrogen layer (at constant effective temperature). As illustrated in Fig.\,\ref{fig:phot}, a thicker hydrogen layer produces stronger absorption in the Lyman continuum ($\lambda < 912$\,\AA), resulting in a lower far-ultraviolet flux and thus a higher near-ultraviolet and optical flux.

\section{Envelope modelling}
\label{sec:env}

\begin{figure}
\includegraphics[width=\columnwidth,clip=true,trim=5.4cm 15.3cm 5.4cm 3.5cm]{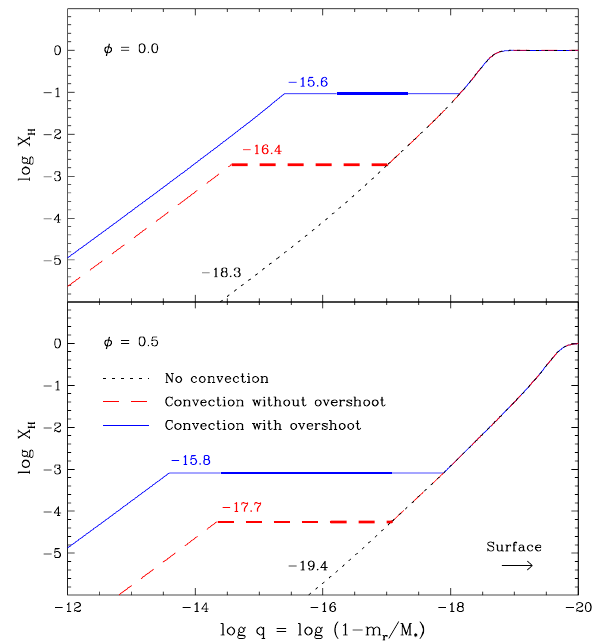}
\caption{Modelled hydrogen abundance profile of ZTF\,J2033 at phases $\phi = 0.0$ (DA face; top panel) and $\phi = 0.5$ (DB face; bottom panel) under various assumptions. The abundance profile is expressed as the run of the hydrogen mass fraction with mass depth $q = 1-m_{r}/M_*$. The dotted black lines show models ignoring convective mixing, while the dashed red and solid blue lines show models including convective mixing without and with overshoot, respectively. In the latter case, overshoot is assumed to extend over two pressure scale heights above and below the formal convection zone, which is indicated by the thicker part of the curve. Each curve is labelled with the respective total mass of hydrogen expressed as $\log M_{\mathrm{H}}/M_*$.}
\label{fig:env}
\end{figure}

Thus far, our analysis has assumed that the radial distribution of hydrogen at a given phase is governed by diffusive equilibrium (i.e., a strict equilibrium between upward gravitational diffusion and downward chemical diffusion; e.g., \citealt{vennes1992b}). This is a useful approximation because the radial composition profile is then entirely determined by a single parameter --- the mass of hydrogen. The hydrogen abundance profiles for $\log M_{\mathrm{H}}/M_* = -18.3$ and $-19.4$, as estimated for ZTF\,J2033 at phases $\phi = 0.0$ and $0.5$, are displayed in Fig.\,\ref{fig:env} as black dotted lines. We use as radial variable the fractional mass depth $q = 1-m_{r}/M_*$ (where $m_r$ is the mass within radius $r$). Although convenient, the diffusive equilibrium approximation is not strictly valid in the presence of convective mixing. At the temperature and gravity of ZTF\,J2033, the helium-rich region just below the hydrogen-rich layer may sustain a small convection zone \citep{rolland2018, cukanovaite2019}, prompting us to refine our estimates of the composition profile and hydrogen mass. This also raises the question of how such thin surface hydrogen layers can remain intact despite the occurrence of convection. Indeed, 3D hydrodynamical simulations of chemically homogeneous envelopes indicate that overshoot beyond the convection zone is very efficient, extending over a few pressure scale heights \citep{kupka2018, cunningham2019, cukanovaite2019}. Simply extrapolating these results to a chemically stratified structure suggests that overshoot may fully dilute the hydrogen layer and produce a well-mixed, helium-dominated atmosphere, apparently at odds with the existence of ZTF\,J2033 and other stratified atmosphere white dwarfs.

To address these questions, we computed 1D static envelope models including the effects of both diffusion and convection using the STELUM code \citep{bedard2022a}. For each phase, we first solved for the thermodynamic structure assuming the composition profile expected from strict diffusive equilibrium. As this initial model contained a small convection zone (treated using the ML2 version of the mixing-length theory; \citealt{tassoul1990}), we then refined the composition profile to incorporate the associated mixing. We recomputed the elemental abundances from the surface inward, integrating the diffusive equilibrium equations in non-convective regions and imposing a uniform composition in convective regions \citep{Koester2020}. Finally, we repeated the above steps, iteratively updating the thermodynamic structure and the composition profile, until convergence to a fully consistent model. With regard to overshoot, we considered two illustrative sets of models: one without any overshoot, and one where overshoot expands the uniformly mixed region by two pressure scale heights above and below the formal convection zone.

The resulting hydrogen abundance profiles for the pure DA and pure DB phases are shown in Fig.\,\ref{fig:env}. The case without overshoot is represented by dashed red lines. As anticipated, these models obey diffusive equilibrium close to the surface but contain a small convection zone (indicated by the thick segment) producing a flat composition profile at $-17.0 \lesssim \log q \lesssim -14.5$. Because some hydrogen is mixed to greater depths, the total mass of hydrogen implied by the observations increases to $\log M_{\mathrm{H}}/M_* = -16.4$ at $\phi = 0.0$ and $\log M_{\mathrm{H}}/M_* = -17.7$ at $\phi = 0.5$. The models including overshoot are depicted as solid blue lines. In this case, the uniformly mixed region is not only larger but also has a higher hydrogen abundance (since it reaches farther into the surface hydrogen layer). Consequently, the inferred total hydrogen masses become even larger, $\log M_{\mathrm{H}}/M_* = -15.6$ and $-15.8$ at $\phi = 0.0$ and $0.5$, respectively.

Our results for the DA face are particularly interesting, as they may hold a clue to the question raised earlier about the stability of a thin hydrogen layer against a putative helium-driven convection zone. With our illustrative overshoot prescription, the mixed region lies relatively close to the overlying pure-hydrogen region and thus has a high hydrogen abundance (mass fraction $X_{\mathrm{H}} \simeq 0.1$). As a result, the formal convection zone is significantly smaller than in the no-overshoot case --- this is the well-known ``poisoning'' effect of hydrogen on helium-driven convection \citep{rolland2018, cukanovaite2019}. With just a little more overshoot (bringing just a little more hydrogen), convection would be suppressed entirely, leading to a physical inconsistency. Therefore, the solution to the puzzle may be that mixing is self-regulating, with overshoot partially diluting the thin hydrogen layer while also being partially inhibited by that very process (an effect that is not considered in existing 3D simulations). This phenomenon where a convective instability is weakened by a stabilising composition gradient is known as semi-convection \citep{shibahashi2007, salaris2017}. Interestingly, ongoing work indicates that semi-convection may occur in a DAQ white dwarf with a stratified hydrogen--carbon atmosphere (Sahu et al., under review). We suggest that the same process could also explain the existence of stratified hydrogen--helium structures. Semi-convection likely produces a mild hydrogen abundance gradient (intermediate between the cases of diffusive equilibrium and uniform mixing), which we however cannot determine here as our models inherently assume that overshoot results in uniform mixing. Although these simple 1D equilibrium models provide rough insights, more complex 3D hydrodynamical simulations are needed to understand the behaviour of overshoot in chemically stratified envelopes.

As noted in Section\,\ref{sec:atmo}, the mixing of trace hydrogen below the surface layer may have an effect on the emergent spectrum, which could potentially be used to constrain the extent of overshoot. To explore this possibility, we performed a simple test where we recomputed the best-fitting synthetic spectra using the same methodology as \citet{bedard2020} but imposing the composition profiles obtained from our envelope models including overshoot. For the DB phase, we find that the spectrum is essentially unchanged, as the hydrogen abundance in the mixed region ($X_{\mathrm{H}} \simeq 0.001$) is too low to affect the atmospheric opacity. For the DA phase, the higher hydrogen content of the mixed region ($X_{\mathrm{H}} \simeq 0.1$) makes the Balmer lines slightly stronger, albeit by less than 5 per cent as this region lies mostly below the line formation depth. Given that mixing is likely less efficient than supposed in this test, as argued above, we conclude that its impact on the spectrum is negligible, justifying the assumption of diffusive equilibrium in our spectroscopic analysis but also ruling out the prospect of constraining overshoot from the observations.

Our various estimates of the hydrogen mass at the surface of ZTF\,J2033 are given in Fig.\,\ref{fig:env}. Note that, interestingly, the overshoot case yields similar values on the DA and DB faces. However, the radial distribution of hydrogen remains different: on the DA side, the hydrogen is concentrated closer to the surface owing to the smaller convection zone. Finally, we caution that the quantity $M_{\mathrm{H}}$ discussed here must be interpreted as the mass of hydrogen in the outer envelope at the present time, and not as the total mass of hydrogen contained in the star. Indeed, a large amount of residual hydrogen is likely still ``hidden'' in the inner envelope and will only emerge later on \citep{rolland2020, bedard2023}.

\section{Discussion and conclusion}
\label{sec:conclu}

Having investigated the current structure of ZTF\,J2033 using static atmosphere and envelope models, we now discuss possible evolutionary interpretations: how has it acquired such a peculiar structure? Given the high effective temperature and overall hydrogen deficiency, \citet{caiazzo2023} suggested that ZTF\,J2033 is undergoing a spectral transformation owing to the diffusion and/or mixing of residual hydrogen. Indeed, the considerations of Section\,\ref{sec:env} confirm that ZTF\,J2033 is at the junction of the two main events in the spectral evolution of hot white dwarfs: a thin surface hydrogen layer is gradually being built by diffusion but is perhaps simultaneously being diluted by nascent convection. Furthermore, the high mass and fast rotation indicate that this object may be the remnant of a stellar merger, which could also explain the origin of the hydrogen deficiency. The outstanding question is, what produces the observed surface asymmetry? \citet{caiazzo2023} proposed two scenarios, both involving a weak asymmetric magnetic field. Although no evidence of magnetism is visible in the spectra, this appears as a reasonable hypothesis given that the upper limit on the field strength is relatively weak ($B \lesssim 1$\,MG) and that other white dwarfs showing composition inhomogeneities are manifestly magnetic \citep{achilleos1992, wesemael2001, kawka2021, bagnulo2024a, bagnulo2024b, cheng2024, moss2024, moss2025}. We review the two scenarios in light of our findings.

The first possibility is that convection is entirely suppressed by the magnetic field --- a plausible assumption as this should occur for $B \gtrsim 10$\,kG \citep{tremblay2015b, cunningham2021, caiazzo2023}. In this case, hydrogen would freely float on the surface (as in our models in strict diffusive equilibrium), and the asymmetry would arise from the influence of the magnetic field on diffusion. Indeed, ionised hydrogen atoms should spiral around (and thus diffuse along) magnetic field lines, ultimately accumulating at the magnetic pole --- this is the so-called ``ocean'' scenario of \citet{caiazzo2023}. Although this argument is qualitatively correct and could still explain our revised structure, the magnitude of the effect is known to be negligible in typical magnetic white dwarfs \citep{michaud1982} and thus deserves a more quantitative estimate.

Because rising hydrogen ions tend to follow magnetic field lines, the vertical component of the diffusion velocity is unchanged where the field is vertical, while it is most reduced where the field is horizontal. The reduction factor is given by (e.g., \citealt{michaud2015})
\begin{equation}
f_B = \frac{1}{1 + \omega_{B}^{2} t_{\mathrm{coll}}^{2}},
\end{equation}
where $\omega_{B} = e B / m_{\mathrm{H}} c$ is the cyclotron frequency (with $e$ the elementary charge, $m_{\mathrm{H}}$ the hydrogen atomic mass, and $c$ the speed of light) and $t_{\mathrm{coll}}$ is the interparticle collision timescale. The latter quantity can be estimated as $t_{\mathrm{coll}} \sim v_{\mathrm{diff},r} / g$, where $v_{\mathrm{diff},r}$ is the (unperturbed) diffusion velocity in the radial direction. In white dwarfs, the velocity is dominated by gravitational diffusion (driven by the pressure gradient) and is thus given by (e.g., \citealt{paquette1986b})
\begin{equation}
v_{\mathrm{diff},r} \sim -D_{\mathrm{H}} \frac{\partial \ln P}{\partial r} = D_{\mathrm{H}} \frac{\rho g}{P},
\end{equation}
where $D_{\mathrm{H}}$ is the diffusion coefficient of hydrogen in helium, $P$ is the pressure, and $\rho$ is the density, and where we have ignored a factor of order unity. Adopting $\log q = -18.5$ as a representative depth and inserting the values from our envelope model of ZTF\,J2033 (using the diffusion coefficient of \citealt{paquette1986a}), we find that $f_B < 0.5$ requires $B \gtrsim 300$\,kG, and $f_B < 0.1$ requires $B \gtrsim 900$\,kG. Therefore, this scenario is possible but implies a magnetic field close to the current upper limit; it could potentially be ruled out if a tighter constraint becomes available.

The second possibility is that convection is inhibited by the magnetic field over part of the stellar surface. In the so-called ``dilution'' scenario of \citet{caiazzo2023}, the magnetic field is high enough on one side that mixing is prevented and hydrogen is able to float, and low enough on the other side that mixing occurs and hydrogen is fully diluted within the convection zone. Of course, this is no longer a viable interpretation as we have demonstrated that hydrogen floats over the full surface. Nevertheless, a variant of this scenario may still be applicable. Our envelope models (specifically those with overshoot, shown as solid blue curves in Fig.\,\ref{fig:env}) indicate that the same amount of hydrogen can be radially distributed in different ways depending on the size of the convection zone, naturally leading to pure-hydrogen layers of different thicknesses. From an evolutionary perspective, only a small asymmetry in the magnetic field (and thus mixing efficiency) may be sufficient for the pure-hydrogen layer to remain thicker on one side for a longer time. This is an attractive idea as it does not require as radical a difference between the two faces as the original suggestion of \citet{caiazzo2023}.

We also wish to draw attention to a third possibility, one that does not necessitate a magnetic field. As discussed in Section\,\ref{sec:env}, it is plausible that convective mixing is inherently impeded by the ``poisoning'' effect of hydrogen on helium-driven convection. Then, the current asymmetry in hydrogen layer thickness may reflect an uneven internal distribution of hydrogen inherited from the past merger, which has been preserved as upward diffusion proceeded. The condition for this scenario to hold is that diffusion in the lateral direction must be slow enough that hydrogen has not been redistributed evenly at the current cooling age of ZTF\,J2033. An order-of-magnitude estimate of the time needed for a lateral composition gradient to be smoothed out is given by $t_{\mathrm{diff},x} \sim d / v_{\mathrm{diff},x}$, where $d$ is a characteristic distance and $v_{\mathrm{diff},x}$ is the lateral diffusion velocity. The latter arises purely from chemical diffusion (driven by the composition gradient) and is thus given by (e.g., \citealt{paquette1986b})
\begin{equation}
v_{\mathrm{diff},x} \sim -D_{\mathrm{H}} \frac{\partial \ln c_{\mathrm{H}}}{\partial x} \sim D_{\mathrm{H}} \frac{\Delta \ln c_{\mathrm{H}}}{d},
\end{equation}
where $c_{\mathrm{H}} = N_{\mathrm{H}} / (N_{\mathrm{H}} + N_{\mathrm{He}})$ is the hydrogen concentration, and where we have replaced the derivative with a finite difference. Assuming $\Delta \ln c_{\mathrm{H}} \sim 1$, $d \sim \pi R_*$, and the diffusion coefficient of our envelope model at $\log q = -18.5$, we obtain $t_{\mathrm{diff},x} \sim 80$\,Myr. This is indeed comparable to the cooling age $t_{\mathrm{cool}} = 145_{\scaleto{-60}{4pt}}^{\scaleto{+65}{4pt}}$\,Myr. This is also much longer than the $\sim$1 Myr duration of the thin hydrogen layer phase, as estimated from theoretical simulations and population statistics of stratified atmosphere white dwarfs \citep{bedard2020, bedard2022a}. Ultimately, 3D (magneto)hydrodynamical models of element transport would be valuable to discriminate between the three possible explanations.

In conclusion, the key result of this Letter is that ZTF\,J2033 has a thin hydrogen layer covering its entire surface, and the photometric and spectroscopic variations arise from changes in layer thickness across the surface. In other words, ZTF\,J2033 may be considered as spectroscopically double-faced, but not truly physically so. Thus, the structure asymmetry is less drastic than initially believed, and the physical interpretation accordingly requires less extreme assumptions. Incidentally, we note that while ZTF\,J2033 has been presented as a two-sided white dwarf, the existing data do not exclude a more common dipolar geometry. For instance, other variable white dwarfs have been modelled with hydrogen polar caps and a helium equatorial belt \citep{pereira2005, moss2024, moss2025}. Similarly, it is entirely possible that ZTF\,J2033 has two spots (rather than one side) where the hydrogen layer is thinner, which would then imply a rotation period of $\simeq$30 (rather than 15) minutes.

Our study has important implications for the potential discovery of other similar objects. ZTF\,J2033 was first found to exhibit a non-uniform surface composition, and then turned out to have a stratified atmosphere. Conversely, could known stratified atmosphere white dwarfs turn out to have non-uniform hydrogen layers? Most of the $\simeq$30 known stratified objects have been identified via a single epoch of Sloan Digital Sky Survey spectroscopy \citep{manseau2016, bedard2020}. Apart from the narrower lines due to lower gravities, their spectra (see fig.\,10 of \citealt{bedard2020}) are remarkably similar to the phase-averaged spectrum of ZTF\,J2033 (see fig.\,2 of \citealt{caiazzo2023}). Monitoring these stars for photometric and spectroscopic variability could prove an efficient way to expand the small class of inhomogeneous white dwarfs. Finally, our work may also help improve our understanding of the current cooler members of this class. At first glance, a stratified configuration may appear unlikely for these stars, as their lower temperatures should result in more vigorous convection. However, the presence of abundance patches on the surface necessarily implies that mixing is somehow impeded. Therefore, the possibility of a vertically (as well as horizontally) inhomogeneous atmosphere should be considered in these cases too.

\section*{Acknowledgements}

We thank Evan Bauer, Pierre Bergeron, Simon Blouin, Ilaria Caiazzo, Falk Herwig, Mike Montgomery, Nicole Reindl, and Olivier Vincent for useful discussions. AB is a Postdoctoral Fellow of the Natural Sciences and Engineering Research Council (NSERC) of Canada. This work was supported by the European Research Council (ERC) under the European Union’s Horizon 2020 research and innovation programme (grant agreement 101002408). The authors acknowledge the University of Warwick's Scientific Computing Research Technology Platform (SCRTP) for assistance in the research described in this paper.

\section*{Data Availability}
 
The data analysed in this work are publicly available in \citet{caiazzo2023}. The atmosphere and envelope models will be shared upon reasonable request.



\bibliographystyle{mnras}
\bibliography{references} 








\bsp	
\label{lastpage}
\end{document}